\documentclass[aps,showkeys,twocolumn,showpacs,preprintnumbers,amsmath,amssymb,superscriptaddress,floatfix,nofootinbib]{revtex4}
\usepackage{graphicx,color,dcolumn,booktabs,bm}
\usepackage{longtable,lscape}
\usepackage{txfonts}
\usepackage{overpic}
\usepackage{epsfig}
\usepackage{amssymb}
\usepackage{rotating}
\usepackage{epstopdf}
\usepackage{indentfirst}
\usepackage{feynmf}   %{feynmp}
\usepackage{slashed}  %for Feynman symbols
\usepackage{cases}
\usepackage{color}
\usepackage{multirow}
\usepackage{graphicx,color,dcolumn,booktabs,bm}
\usepackage{cases}
\usepackage{array}

\begin{document}

\title{Mass spectra and radiative transitions of doubly heavy baryons in a relativized quark model}

\author{Qi-Fang L\"{u}} \email{lvqifang@ihep.ac.cn}
\affiliation{Department of Physics, Hunan Normal University, and Key Laboratory of Low-Dimensional Quantum
Structures and Quantum Control of Ministry of Education, Changsha 410081, China}
\affiliation{Synergetic Innovation Center for Quantum Effects and Applications (SICQEA),
Hunan Normal University, Changsha 410081, China}
\author{Kai-Lei Wang}
\affiliation{Department of Physics, Hunan Normal University, and Key Laboratory of Low-Dimensional Quantum
Structures and Quantum Control of Ministry of Education, Changsha 410081, China}
\affiliation{Synergetic Innovation Center for Quantum Effects and Applications (SICQEA),
Hunan Normal University, Changsha 410081, China}
\author{Li-Ye Xiao} %\email{lyxiao@pku.edu.cn}
\affiliation{School of Physics and State Key Laboratory of Nuclear Physics and Technology, Peking University, Beijing 100871, China}
\author{Xian-Hui Zhong} \email{zhongxh@hunnu.edu.cn}
\affiliation{Department of Physics, Hunan Normal University, and Key Laboratory of Low-Dimensional Quantum
Structures and Quantum Control of Ministry of Education, Changsha 410081, China}
\affiliation{Synergetic Innovation Center for Quantum Effects and Applications (SICQEA),
Hunan Normal University, Changsha 410081, China}

\begin{abstract}

We study the mass spectra and radiative decays of doubly heavy baryons within the diquark picture in a relativized quark model.
The mass of the $J^P=1/2^+$ $\Xi_{cc}$ ground state is predicted to be 3606 MeV, which is consistent with the mass of $\Xi_{cc}^{++}(3621)$ newly observed by the LHCb collaboration. The predicted mass gap between two $S$ wave states, $\Xi_{cc}^*$ ($J^P=3/2^+$) and $\Xi_{cc}$ ($J^P=1/2^+$), is 69 MeV. Furthermore, the radiative transitions of doubly heavy baryons are also estimated by using the realistic wave functions obtained from relativized quark model. The radiative decay widths of $\Xi_{cc}^{*++} \to \Xi_{cc}^{++}\gamma$ and $\Xi_{cc}^{*+} \to \Xi_{cc}^+\gamma$ are predicted to be about 7 and 4 keV, respectively. These predictions of doubly heavy baryons can provide helpful information for future experimental searches.

\end{abstract}


\maketitle

\section{Introduction}{\label{introduction}}

The doubly heavy baryons, made up of two heavy quarks and one light quark, are particularly
interesting because they provide a new platform for studying the heavy quark symmetry and chiral dynamics simultaneously~\cite{Richard:1992uk,Klempt:2009pi,Chen:2016spr}.
To looking for the doubly heavy baryons, great efforts were made
in the past two decades. In 2002, the SELEX collaboration reported some
evidence of a doubly charmed baryon $\Xi_{cc}^+(3519)$ in the $\Lambda_c^+ K^- \pi^+$
final state, and confirmed it in the $pD^+K^-$ decay mode~\cite{Mattson:2002vu,Ocherashvili:2004hi}.
However, $\Xi_{cc}^+(3519)$ cannot be confirmed by the FOCUS, BaBar, Belle and LHCb collaborations~\cite{Ratti:2003ez,Aubert:2006qw,Chistov:2006zj,Aaij:2013voa}.
Very recently, the LHCb collaboration observed a highly significant structure with a mass
of $M\simeq 3621$ MeV in the $\Lambda_c^+ K^- \pi^+ \pi^-$ mass spectrum in the proton-proton collisions~\cite{Aaij:2017ueg}.
This structure can be identified as a doubly charmed baryon (denoted by $\Xi_{cc}^{++}(3621)$) from its weakly decaying behaviors. Due to the large mass difference, the newly observed $\Xi_{cc}^{++}(3621)$ by the LHCb collaboration and $\Xi_{cc}^+(3519)$ reported by the SELEX collaboration can be hardly categorized as a isodoublet~\cite{Aaij:2017ueg,Karliner:2017gml}.

In theory, the mass spectra of the doubly heavy baryons have been predicted with various methods, such as the constituent quark model~\cite{Kiselev:2001fw,Ebert:1996ec,Ebert:2002ig,Gershtein:2000nx,Roberts:2007ni,Giannuzzi:2009gh,Martynenko:2007je,Valcarce:2008dr,Shah:2017liu}, heavy quark symmetry and mass formulas~\cite{Roncaglia:1995az,Cohen:2006jg,Eakins:2012jk,Karliner:2014gca}, Regge behaviors~\cite{Wei:2015gsa,Wei:2016jyk}, QCD sum rule~\cite{Zhang:2008rt,Tang:2011fv,Wang:2010hs,Aliev:2012ru,Aliev:2012nn,Aliev:2012iv}, lattice QCD~\cite{Liu:2009jc,Brown:2014ena,Padmanath:2015jea} and so on. For the lowest $\Xi_{cc}$ state with $J^P=1/2^+$, the theoretical predicted masses lie in a large range $3500\sim 3700$ MeV, thus, both $\Xi_{cc}^{++}(3621)$ and $\Xi_{cc}^+(3519)$ can be candidates of the ground $\Xi_{cc}$ state with $J^P=1/2^+$. To clarify which one should be the lowest $\Xi_{cc}$ state, further investigations are needed. Besides the mass spectra, the weak decays of the doubly heavy baryons are also extensively discussed in the literature~\cite{Faessler:2001mr,Faessler:2009xn,Albertus:2009ww,White:1991hz,Li:2017ndo,Yu:2017zst,Ebert:2004ck,Roberts:2008wq}.
However, the studies on the radiative decays are scarce, only several works focusing on the ground doubly heavy baryons are found in the literature~\cite{Branz:2010pq,Hackman:1977am,Bernotas:2013eia,Dai:2000hza,Albertus:2010hi,Hu:2005gf}.
A comprehensive review on the doubly heavy baryons can be found in Ref.~\cite{Chen:2016spr}.

Lately, stimulated by the newly observed state $\Xi_{cc}^{++}(3621)$ by the LHCb collaboration, several theoretical works were performed. Using the QCD sum rule, Chen \emph{et al.} investigated the low-lying doubly charmed baryon spectra, which suggests the newly observed $\Xi_{cc}^{++}(3621)$ may be the ground $1/2^+$ state~\cite{Chen:2017sbg}. The masses and wave functions of ground doubly charmed baryons were also restudied with the Cornell potential~\cite{Kerbikov:2017pau}.  The radiative transitions between the ground doubly charmed baryons were investigated within the heavy baryon chiral perturbation theory~\cite{Li:2017pxa} and constituent quark model with simple harmonic oscillator wave functions~\cite{Xiao:2017udy}. Furthermore, the magnetic moments, weak decays, and other related topics are also discussed in the literature ~\cite{Li:2017cfz,Wang:2017mqp,Wang:2017azm,Meng:2017udf,Karliner:2017qjm,Eichten:2017ffp,
Brambilla:2017uyf,Gutsche:2017hux,Karliner:2017elp,Guo:2017vcf}.

In this work, we use a relativized quark model to calculate the mass spectra of doubly heavy baryons. The relativized quark model, proposed by Godfrey, Capstick, and Isgur, has been extensively used to study the properties of the conventional hadrons and gives a unified description of the hadron spectra~\cite{Godfrey:1985xj,Capstick:1986bm,Godfrey:1998pd,Godfrey:2015dia,Ferretti:2013faa, Ferretti:2013vua,Lu:2014zua,Godfrey:2015dva,Song:2015nia}. Therefore, it is suitable to deal with the doubly heavy baryons, where both heavy-heavy and heavy-light systems are included. Moreover, the relativistic effects are also involved in this model, which may be essential for the light quarks. In Ref.~\cite{Capstick:1986bm}, Capstick and Isgur adopted this relativized quark model to estimate various baryon spectra, however, they did not extend it to deal with the doubly heavy baryon spectra. Given the similarity between the heavy diquark and heavy antiquark, the doubly heavy baryons looks like heavy-light mesons in the diquark picture~\cite{Ebert:1996ec,Ebert:2002ig,Ebert:2007rn,Ebert:2008kb,Ebert:2010af,
Monemzadeh:2014kra,Hadizadeh:2015cvx,Ferretti:2011zz,Santopinto:2014opa,Lu:2016cwr,Lu:2016zhe}.
Within the diquark picture, we first calculate the masses and wave functions of the $cc$ and $bb$ diquarks.
Then, the mass spectra of the doubly heavy baryons and the diquark-quark wave functions can be
obtained by solving the Schr\"{o}dinger-type
equation between the diquark and quark. Finally, the total wave function of a doubly heavy baryon can
be expressed as the diquark wave function multiplied by the diquark-quark wave function.
It should be pointed out that when we treat the diquark-quark interactions,
the diquark size effects to the potentials are also considered by introducing
a form factor as that adopted in Ref.~\cite{Ebert:2002ig}. Our predicted mass for the ground doubly charmed baryon with $J^P=1/2^+$
is 3606 MeV. It suggests that the newly observed $\Xi_{cc}^{++}(3621)$ state
may be assigned as the $J^P=1/2^+$ ground state.

Using the wave functions obtained from the relativized quark model, we
further study the radiative transitions of doubly heavy baryons.
In the calculations, we emplore an EM transition operator which is extracted
in the non-relativistic constituent quark model and has been successfully applied
to the study of the radiative decays of $c\bar{c}$ and $b\bar{b}$ systems
~\cite{Deng:2016stx,Deng:2016ktl} and the heavy baryons~\cite{Wang:2017hej,Xiao:2017udy}.
Due to the absence of hadronic transitions for the low-lying states~\cite{Xiao:2017udy},
these electromagnetic decays provide useful information of the internal structures.
It is found that the partial decay widths for $\Xi_{cc}^{*++} \to \Xi_{cc}^{++}\gamma$
and $\Xi_{cc}^{*+} \to \Xi_{cc}^+\gamma$ are predicted to be about 7 and 4 keV,
respectively, which is quite significant and helpful for future experimental searches.

This paper is organized as follows. The relativized quark model is briefly introduced, and then the masses of the heavy diquarks and doubly heavy baryons are presented in Sec.~\ref{spectra}. The radiative decays of doubly heavy baryons are estimated in Sec.~\ref{transitions}. A short summary is given in the last section.

\begin{table}[!htbp]
\begin{center}
\caption{ \label{tab1} The masses and relevant form factor parameters of the $cc$ and $bb$ diquarks. The notation $n^{2S+1}L_{J}$ is used to stand for the ground and excited diquarks. The brace corresponds to symmetric quark content in flavor.}
\normalsize
\begin{tabular*}{8.5cm}{@{\extracolsep{\fill}}*{5}{p{1.5cm}<{\centering}}}
\hline\hline
 Quark content &  Diquark type & Mass (MeV)     &  $\xi$ (GeV)      &  $\zeta$ (\rm{$GeV^2$}) \\\hline
 $\{cc\}$     &  $1^3S_1$            & 3294    &  1.535    &   0.245\\
 $\{cc\}$     &  $1^1P_1$            & 3507    &  0.673    &   0.367\\
 $\{cc\}$     &  $2^3S_1$            & 3603    &  0.714    &   0.123\\
 $\{cc\}$     &  $1^3D_1$            & 3667    &  0.429    &   0.299\\
 $\{bb\}$     &  $1^3S_1$            & 9823    &  2.063    &   1.010\\
 $\{bb\}$     &  $1^1P_1$            & 10016   &  0.874    &   0.773\\
 $\{bb\}$     &  $2^3S_1$            & 10088   &  0.879    &   0.298\\
 $\{bb\}$     &  $1^3D_1$            & 10144   &  0.547    &   0.532\\
 \hline\hline
\end{tabular*}
\end{center}
\end{table}

\begin{table*}[!htbp]
\begin{center}
\caption{ \label{mass1} Masses of ground states of the doubly heavy baryons compared with different calculations. The units are in MeV.}
\normalsize
\begin{tabular*}{17.5cm}{@{\extracolsep{\fill}}*{9}{p{1.65cm}<{\centering}}}
\hline\hline
 Baryon          &  Content            & $J^P$       & Our work   & RQM~\cite{Ebert:2002ig} & NQM~\cite{Gershtein:2000nx}     & FH~\cite{Roncaglia:1995az}  &  NQM~\cite{Roberts:2007ni}  & LQCD~\cite{Brown:2014ena} \\\hline
 $\Xi_{cc}$      &  $\{cc\}q$          & $1/2^+$     & 3606       & 3620                 & 3478                         &  3660          & 3676  & 3610   \\
 $\Xi_{cc}^*$    &  $\{cc\}q$          & $3/2^+$     & 3675       & 3727                 & 3610                         &  3740          & 3753  & 3692\\
 $\Omega_{cc}$   &  $\{cc\}s$          & $1/2^+$     & 3715       & 3778                 & 3590                         &  3740          & 3815  & 3738 \\
 $\Omega_{cc}^*$ &  $\{cc\}s$          & $3/2^+$     & 3772       & 3872                 & 3690                         &  3820          & 3876 & 3822 \\
 $\Xi_{bb}$      &  $\{bb\}q$          & $1/2^+$     & 10138      & 10202                & 10093                        &  10340          & 10340 &10143  \\
 $\Xi_{bb}^*$    &  $\{bb\}q$          & $3/2^+$     & 10169      & 10237                & 10133                        &  10370          & 10367  &10178  \\
 $\Omega_{bb}$   &  $\{bb\}s$          & $1/2^+$     & 10230      & 10359                & 10180                        &  10370          & 10454  &10273   \\
 $\Omega_{bb}^*$ &  $\{bb\}s$          & $3/2^+$     & 10258      & 10389                & 10200                        &  10400          & 10486  & 10308 \\

\hline\hline
\end{tabular*}
\end{center}
\end{table*}

\section{Mass spectra}{\label{spectra}}

The Hamiltonian between quark and antiquark in the relativized quark model can be expressed as
\begin{equation}
\tilde{H} = H_0+\tilde{V}(\boldsymbol{p},\boldsymbol{r}), \label{ham}
\end{equation}
\begin{equation}
H_0 = (p^2+m_i^2)^{1/2}+(p^2+m_j^2)^{1/2},
\end{equation}
\begin{equation}
\tilde{V}(\boldsymbol{p},\boldsymbol{r}) = \tilde{H}^{conf}_{ij}+\tilde{H}^{cont}_{ij}+\tilde{H}^{ten}_{ij}+\tilde{H}^{so}_{ij},
\end{equation}
where the $\tilde{H}^{conf}_{ij}$ includes the spin-independent linear confinement and Coulomb-like interaction, the $\tilde{H}^{cont}_{ij}$ is the color contact term, the $\tilde{H}^{ten}_{ij}$ is the color tensor interaction, and $\tilde{H}^{so}_{ij}$ is the spin-orbit term. $\boldsymbol{p}$ and $\boldsymbol{r}$ stand for the relative momentum and coordinate between the quark and antiquark. $\tilde{H}$ denotes an operator that has taken account of the relativistic effects according to the relativized procedure. The explicit forms of these interactions and the details of this relativization scheme can be found in Ref.~\cite{Godfrey:1985xj,Capstick:1986bm}. In the baryons, only the $\bar 3$ type diquarks in the color space exist, hence the relation $\tilde{V}_{qq}(\boldsymbol{p},\boldsymbol{r})=\tilde{V}_{q\bar q}(\boldsymbol{p},\boldsymbol{r})/2$ is employed. Since the original parameters of Ref.~\cite{Capstick:1986bm} can describe the low lying baryon spectra well, we use the same parameters to investigate the doubly heavy baryons. Moreover, for the two-body interaction constant, $c=-253~\rm{MeV}$ is adopted as the same in Refs.~\cite{Godfrey:1985xj,Capstick:1986bm}.

\begin{figure}[!htbp]
\includegraphics[scale=0.7]{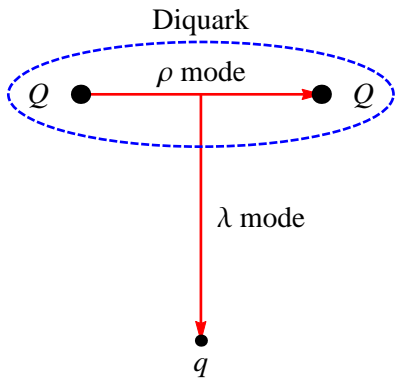}
\vspace{0.0cm} \caption{Doubly heavy baryon system with Jacobi coordinates defined as $\rho=\frac{1}{\sqrt{2}}(\boldsymbol{r_1}-\boldsymbol{r_2})$ and $\lambda=\frac{1}{\sqrt{6}}(\boldsymbol{r_1}+\boldsymbol{r_2}-2\boldsymbol{r_3})$. $Q$ and $q$ stand for the heavy quark and light quark, respectively.}
\label{coordinate}
\end{figure}

The structures of the doubly heavy baryons are  illustrated in Fig.~\ref{coordinate}, where the two heavy quarks are treated as a diquark with a size. Here, the $\rho$-mode excitaion corresponds to the internal excitation of the diquark, and the $\lambda$-mode excitation stands for the diquark-quark excitation. When the $cc$ and $bb$ diquarks locate in $1S$ wave, the spin-parities of the diquarks are $J^P=0^+$ and $J^P=1^+$, named as the scalar diquarks and axial diquarks, respectively. Constrained by the symmetry, the scalar diquark cannot exist in the $cc$ and $bb$ systems. In the doubly heavy baryons, the first excited state comes from the diquark ($\rho$-mode excitation) rather than the light quark ($\lambda$-mode excitation), so we have to discuss the excited diquarks as well. Here, we only consider the $1^1P_1$, $2^3S_1$, and $1^3D_1$ diquarks in the low lying baryons. We use the Gaussian expansion method to solve the Hamiltonian~(\ref{ham}) with $\tilde{V}_{qq}(\boldsymbol{p},\boldsymbol{r})$ potential~\cite{Hiyama:2003cu}. The obtained masses of these diquarks are presented in Table.~\ref{tab1}. The form factors $F(r)$ are introduced to reflect the diquark sizes. The $F(r)/r$ can be obtained from the Fourier transforms of $F(\boldsymbol{k}^2)/\boldsymbol{k}^2$, and $F(\boldsymbol{k}^2)$ can be taken as~\cite{Ebert:2002ig}
\begin{equation}
F(\boldsymbol{k}^2) = \frac{\sqrt{E_dM_d}}{E_d+M_d}\Bigg[ \int \frac{d^3 \boldsymbol p}{(2\pi)^3}\bar \Psi_d (\boldsymbol p + \frac{2m_{Q_2}}{E_d+M_d} \boldsymbol k) \Psi_d(\boldsymbol p) + (1 \leftrightarrow 2)  \Bigg],
\end{equation}
where the $E_d$, $M_d$, $\boldsymbol k$, $\Psi_d$ are the energy, mass, total momentum, and wave function of the diquark, respectively.
The form factors can be further approximated by the following expression~\cite{Ebert:2002ig}
\begin{equation}
F(r) = 1-e^{-\xi r -\zeta r^2},
\end{equation}
where the $\xi$ and $\zeta$ are the real numbers. The relevant parameters are calculated by using the obtained diquark wave functions, and also listed in Table.~\ref{tab1}.

With the diquarks listed in Table.~\ref{tab1}, one can calculate the masses  of the doubly heavy baryons and the wave functions between diquarks and quarks. Then, the total wave function of the doubly heavy baryon can be expressed as the wave function of diquark ($\rho$ mode) multiplied by the wave function of diquark-quark ($\lambda$ mode).  The masses of ground states of the doubly heavy baryons together with different calculations are presented in Tab.~\ref{mass1}. The predicted mass of $J^P=1/2^+$ $\Xi_{cc}$ ground state is 3606 MeV, which is consistent with the newly observed $\Xi_{cc}^{++}(3621)$ by the LHCb collaboration. This assignment also agrees with many other works~\cite{Ebert:2002ig,Roberts:2007ni,Brown:2014ena,Chen:2017sbg,Kerbikov:2017pau}. Our results indicate that the $\Xi_{cc}^+(3519)$ reported by SELEX collaboration may not be interpreted as the conventional $\Xi_{cc}$ state, and more experimental data are needed to clarify its nature.

In the diquark picture, the $\{cc\}q$ configuration is similar to a heavy-light meson $\bar c q$ and $\bar b q$, where the heavy diquark acts as a heavy antiquark with size. Experimentally, the spin splittings of the heavy-light mesons decrease when the heavy quark mass increases. In our calculation, the axial ${cc}$ diquark mass lies between the masses of $c$ and $b$ quarks, hence the $S$-wave mass gap should satisfy the relation
\begin{equation}
m(B^*)-m(B) \leq m(\Xi_{cc}^*) - m(\Xi_{cc}) \leq m(D^*) - m(D). \label{cons}
\end{equation}
Our predicted mass of $\Xi_{cc}^*$ is 3675 MeV, and the mass gap between the two $S$-wave $\Xi_{cc}$ states is about 69 MeV.
Although all the predictions in the Tab.~\ref{mass1} satisfy the above relation, the mass gaps from different
models vary from 69 to 132 MeV. It should be pointed out that the nonrelativistic potential model gives rather
small masses for the ground states~\cite{Gershtein:2000nx}. These divergences may help to test different models
in future experiments.

\begin{figure}[!htbp]
\includegraphics[scale=0.45]{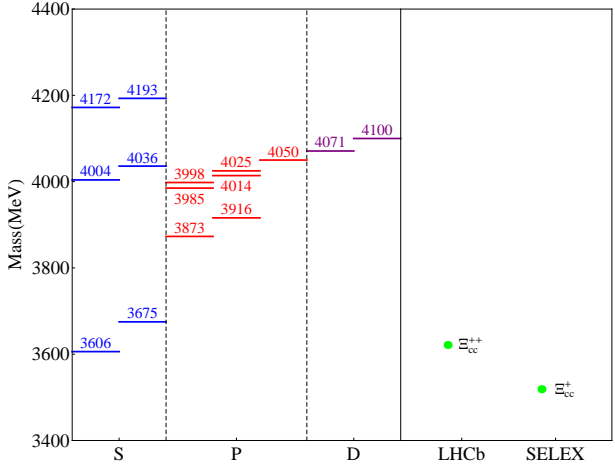}
\vspace{0.0cm} \caption{The predicted mass spectra of the $\Xi_{cc}$ baryons.}
\label{mass}
\end{figure}

Furthermore, our predicted mass spectra of $\Xi_{cc}$, $\Omega_{cc}$, $\Xi_{bb}$, and $\Omega_{bb}$ are given in Tab.~\ref{mass2}, and the $\Xi_{cc}$ spectra together with the experimental data are also shown in Fig.~\ref{mass}. In this work, we adopt the notation $(N_dL_dn_ql_q)J^P$ to stand for the doubly heavy baryon states, where $N_d$, $L_d$, $n_q$, and $l_q$ correspond to the quantum numbers of the diquark internal radial excitation, diquark internal orbital excitation, light quark radial excitation and light quark orbital excitation, respectively. For the $(1S1p)1/2^-$  and $(1S1p)3/2^-$ states, the additional total spin $S_T$ is introduced to distinguish the spin singlet and triplet.  From Tab.~\ref{mass2}, we can see that the  $P$-wave diquark excitations are the lowest excited states, while the  $2S$ diqaurk, $1D$ diqaurk, and $P$-wave light quark excitations lie in the same energy region. It should be emphasized that for the doubly heavy baryons, the heavy diquark is more easily excited due to the its larger reduced mass, however, for the singly heavy baryons, the light diquark is more difficultly excited due to its small reduced mass, and the excitation mainly exists between the heavy quark and light diquark.

\begin{table}[!htbp]
\begin{center}
\caption{ \label{mass2} Predicted mass spectra of $\Xi_{cc}$, $\Omega_{cc}$, $\Xi_{bb}$, and $\Omega_{bb}$. The units are in MeV.}
\normalsize
\begin{tabular}{p{2.5cm}<{\centering}p{1.3cm}<{\centering}p{1.3cm}<{\centering}p{1.3cm}<{\centering}p{1.3cm}<{\centering}}
\hline\hline
$(N_dL_dn_ql_q)J^P$       & $\Xi_{cc}$  & $\Omega_{cc}$  &  $\Xi_{bb}$  & $\Omega_{bb}$ \\\hline
 $(1S1s)1/2^+$        & 3606        &  3715          & 10138  & 10230          \\
 $(1S1s)3/2^+$        & 3675        &  3772          & 10169  & 10258         \\
 $(1S1p)1/2^-~_{S_T=1/2}$      & 3998        &  4087          & 10525  & 10605        \\
 $(1S1p)3/2^-~_{S_T=1/2}$        & 4014        &  4107          & 10526  & 10610      \\
 $(1S1p)1/2^-~_{S_T=3/2}$         & 3985        &  4081          & 10504  & 10591       \\
 $(1S1p)3/2^-~_{S_T=3/2}$        & 4025        &  4114          & 10528  & 10611       \\
 $(1S1p)5/2^-$         & 4050        &  4134          & 10547  & 10625        \\
 $(1S2s)1/2^+$       & 4172        &  4270          & 10662  & 10751      \\
 $(1S2s)3/2^+$          & 4193        &  4288          & 10675  & 10763 \\\hline
 $(1P1s)1/2^-$         & 3873        &  3986          & 10364  & 10464 \\
 $(1P1s)3/2^-$       & 3916        &  4020          & 10387  & 10482 \\\hline
 $(2S1s)1/2^+$         & 4004        &  4118          & 10464  & 10566 \\
 $(2S1s)3/2^+$         & 4036        &  4142          & 10480  & 10579 \\\hline
  $(1D1s)1/2^+$         & 4071        &  4186          & 10522  & 10625  \\
 $(1D1s)3/2^+$         & 4100        &  4207          & 10538  & 10638 \\
\hline\hline
\end{tabular}
\end{center}
\end{table}

\section{Radiative transitions}{\label{transitions}}

Besides the mass spectra, the decay behaviors are also needed to search for these doubly heavy baryons in experiments.
Due to the suppression of phase space of hadronic transitions, the radiative transitions should play essential roles
in the decays of the low-lying doubly heavy baryons. To treat one-photon radiative decay of a hadron we apply
an EM transition operator which has been successfully applied to study
the radiative decays of $c\bar{c}$ and $b\bar{b}$ systems~\cite{Deng:2016stx,Deng:2016ktl}
and $\Omega_c$ baryons~\cite{Wang:2017hej}.
In this model, the quark-photon EM coupling at the tree level
is adopted as
\begin{equation}
H_e = -\sum_j e_j \bar \psi_j \gamma_\mu^j A^\mu(\boldsymbol{k},\boldsymbol{r}_j) \psi_j, \label{he1}
\end{equation}
where $\psi_j$ stands for the $j$th quark field with coordinate $\boldsymbol{r}_j$ and $A^\mu$ is the photon field with three-momentum $\boldsymbol{k}$. To match the wave functions obtained by the Schr\"{o}dinger-type equation,
we adopt the quark-photon EM couplings in a nonrelativistic form. In the initial-hadron-rest system,
the approximate form can be written as~\cite{Deng:2016stx,Deng:2016ktl,Brodsky:1968ea,Li:1997gd,Zhao:2002id,Xiao:2015gra,Zhong:2011ti,Zhong:2011ht}
\begin{equation}
h_e \cong \sum_j \Bigg [ e_j \boldsymbol{r}_j \cdot \boldsymbol \epsilon - \frac{e_j}{2m_j} \boldsymbol \sigma_j \cdot (\boldsymbol \epsilon \times \boldsymbol {\hat k}) \Bigg ] e^{-i\boldsymbol{k} \cdot \boldsymbol{r}_j}, \label{he2}
\end{equation}
where $e_j$, $m_j$, and $\boldsymbol \sigma_j$ stand for the charge, consistent mass, Pauli spin vector for the $j$th quark, respectively. The $\boldsymbol \epsilon$ is the polarization vector of the final photon.

One can obtain the standard helicity amplitude $\mathcal{A}$ of the radiative process~\cite{Deng:2016stx,Deng:2016ktl}
\begin{equation}
\mathcal{A} = -i \sqrt{\frac{\omega_\gamma}{2}} \langle f | h_e | i \rangle.
\end{equation}
Then, we can estimate the radiative transitions straightforward~\cite{Deng:2016stx,Deng:2016ktl}
\begin{equation}
\Gamma = \frac{|\boldsymbol k|^2}{\pi} \frac{2}{2J_i+1} \frac{M_f}{M_i} \sum_{J_{fz},J_{iz}} |\mathcal{A}|^2,
\end{equation}
where $J_i$ is the total angular momentum of the initial baryons, and $J_{fz}$ and $J_{fi}$ are the components of the total angular momenta along the $z$ axis of the initial and final baryons, respectively. In present calculation, the masses and wave functions of doubly heavy baryons are adopted from our theoretical predictions.

The radiative transitions between the $J^P=3/2^+$ and $J^P=1/2^+$ ground doubly charmed and bottom baryons are predicted, and the results together with different calculations are listed in Tab.~\ref{decay1}. For the $J^P=1/2^+$ ground state, only weak decays can occur. Since the strong decays are forbidden, the electromagnetic transitions dominate for the  $J^P=3/2^+$ ground state. The theoretical radiative decay widths for the $\Xi_{cc}^{*++}\to \Xi_{cc}^{++} \gamma$ and $\Xi_{cc}^{*+}\to\Xi_{cc}^{+~} \gamma$ transitions are
\begin{eqnarray}
\Gamma[\Xi_{cc}^{*++} &\to & \Xi_{cc}^{++} \gamma] = 7.21~\rm{keV},\\
\Gamma[\Xi_{cc}^{*+~} &\to & \Xi_{cc}^{+~} \gamma] = 3.90~\rm{keV},
\end{eqnarray}
respectively, which are roughly compatible with the quark model predictions~\cite{Branz:2010pq}, simple harmonic oscillator~\cite{Xiao:2017udy},
the bag model~\cite{Hackman:1977am,Bernotas:2013eia}, and chiral perturbation theory calculations~\cite{Li:2017pxa} in magnitude. However, the partial width ratio
\begin{equation}
\frac{\Gamma[\Xi_{cc}^{*++} \to \Xi_{cc}^{++} \gamma]}{\Gamma[\Xi_{cc}^{*+} \to \Xi_{cc}^+ \gamma]} = 1.85,
\end{equation}
predicted by us shows special feature compared to other works~\cite{Branz:2010pq,Xiao:2017udy,Hackman:1977am,Bernotas:2013eia,Li:2017pxa}.
More studies of the radiative decay processes $\Xi_{cc}^{*++}\to \Xi_{cc}^{++} \gamma$ and $\Xi_{cc}^{*+}\to\Xi_{cc}^{+~} \gamma$
are needed in the future. Since the LHCb collaboration has observed $\Xi_{cc}^{++}$ in the $\Lambda_c^+ K^- \pi^+ \pi^-$ mass spectrum,
the large decay rate of $\Xi_{cc}^{*++}\to \Xi_{cc}^{++} \gamma$ indicates that they may establish $\Xi_{cc}^{*++}$
in the $\Lambda_c^+ K^- \pi^+ \pi^-\gamma$
final state via the decay chain of $\Xi_{cc}^{*++}\to \Xi_{cc}^{++}\gamma \to \Lambda_c^+ K^- \pi^+ \pi^-\gamma$.
The radiative partial decay widths of $\Omega_{cc}^*\to \Omega_{cc}\gamma$, $\Xi_{bb}^{*0}\to \Xi_{bb}^{0}\gamma$,
$\Xi_{bb}^{*-}\to \Xi_{bb}^{-}\gamma$, and $\Omega_{bb}^*\to \Omega_{bb}\gamma$ are sensitive
to the masses of the initial and final states. In present work, we predict sizeable partial decay widths
for these processes. These information should be helpful for searching for
the missing $J^P=3/2^+$ ground states in future experiments.

\begin{table*}[!htbp]
\begin{center}
\caption{ \label{decay1} Radiatvie decay widths between the $J^P=3/2^+$ and $J^P=1/2^+$
ground doubly charmed and bottom baryons. The units are in keV.}
\normalsize
\begin{tabular*}{17.5cm}{@{\extracolsep{\fill}}*{7}{p{2.0cm}<{\centering}}}
\hline\hline
  Transition                                        & Our work    & RQM\cite{Branz:2010pq} & SHO\cite{Xiao:2017udy} & BG\cite{Hackman:1977am}     & BG\cite{Bernotas:2013eia}  & CPT\cite{Li:2017pxa}  \\\hline
 $\Xi_{cc}^{*++} \to \Xi_{cc}^{++} \gamma$    & 7.21       & 23.46        & 16.7              & 4.35                         & 1.43   & 22.0    \\
 $\Xi_{cc}^{*+} \to \Xi_{cc}^+ \gamma$        & 3.90       & 28.79         & 14.6             & 3.96                         & 2.08   & 9.57   \\
 $\Omega_{cc}^* \to \Omega_{cc} \gamma$       & 0.82        & 2.11         & 6.93             & 1.35                         & 0.95   & 9.45    \\
 $\Xi_{bb}^{*0} \to \Xi_{bb}^0 \gamma$        & 0.98        & 0.31         &1.19            & $\cdot\cdot\cdot$  & $\cdot\cdot\cdot$ & $\cdot\cdot\cdot$       \\
 $\Xi_{bb}^{*-} \to \Xi_{bb}^- \gamma$        & 0.21        & 0.06         &0.24            & $\cdot\cdot\cdot$  & $\cdot\cdot\cdot$ & $\cdot\cdot\cdot$       \\
 $\Omega_{bb}^* \to \Omega_{bb} \gamma$       & 0.04        & 0.02         &0.08             & $\cdot\cdot\cdot$ & $\cdot\cdot\cdot$ & $\cdot\cdot\cdot$      \\

\hline\hline
\end{tabular*}
\end{center}
\end{table*}

\begin{table*}[!htbp]
\begin{center}
\caption{ \label{decay2} Partial decay widths of the radiative transitions of low-lying excited $\Xi_{cc}^{++}$, $\Xi_{cc}^+$, $\Omega_{cc}$, $\Xi_{bb}^0$, $\Xi_{bb}^-$, and $\Omega_{bb}$ states. The units are in keV.}
\normalsize
\begin{tabular}{p{6cm}<{\centering}p{2cm}<{\centering}p{1.6cm}<{\centering}p{1.6cm}<{\centering}p{1.6cm}<{\centering}p{1.6cm}<{\centering}p{1.6cm}<{\centering}}
\hline\hline
  Transition        & $\Xi_{cc}^{++}$   &  $\Xi_{cc}^+$  & $\Omega_{cc}$  & $\Xi_{bb}^0$  &$\Xi_{bb}^-$  & $\Omega_{bb}$ \\\hline
  $|(1P1s)1/2^-\rangle \to |(1S1s)1/2^+ \rangle \gamma$      & 0.85    &0.85            &0.90      &$\sim$0      &$\sim$0       &$\sim$0           \\
 $|(1P1s)3/2^-\rangle \to |(1S1s)1/2^+ \rangle \gamma$      & 1.73      &1.73            &1.57      &0.01      &0.01       &0.01           \\
 $|(1P1s)1/2^-\rangle \to |(1S1s)3/2^+ \rangle \gamma$      & 0.39       &0.39            &0.57      &$\sim$0      &$\sim$0         &$\sim$0           \\
 $|(1P1s)3/2^-\rangle \to |(1S1s)3/2^+ \rangle \gamma$      & 1.03      &1.03            &1.16      &0.01      &0.01         &0.01           \\\hline
 $|(1D1s)1/2^+\rangle \to |(1P1s)1/2^- \rangle \gamma$      & 0.22      &0.22            &0.22      &$\sim$0      &$\sim$0         &$\sim$0           \\
 $|(1D1s)1/2^+\rangle \to |(1P1s)3/2^- \rangle \gamma$      & 0.04        &0.04         &0.05      &$\sim$0      &$\sim$0         &$\sim$0           \\
 $|(1D1s)3/2^+\rangle \to |(1P1s)3/2^- \rangle \gamma$      & 0.12       &0.12            &0.11      &$\sim$0      &$\sim$0         &$\sim$0           \\
 $|(1D1s)3/2^+\rangle \to |(1P1s)3/2^- \rangle \gamma$      & 0.20      &0.20            &0.22      &$\sim$0      &$\sim$0         &$\sim$0           \\\hline
 $|(2S1s)1/2^+\rangle \to |(1P1s)1/2^- \rangle \gamma$      & 0.03       &0.03           &0.03       &$\sim$0      &$\sim$0         &$\sim$0           \\
 $|(2S1s)1/2^+\rangle \to |(1P1s)3/2^- \rangle \gamma$      & 0.01        &0.01            &0.02      &$\sim$0      &$\sim$0         &$\sim$0           \\
 $|(2S1s)3/2^+\rangle \to |(1P1s)1/2^- \rangle \gamma$      & 0.15       &0.15           &0.12      &$\sim$0      &$\sim$0         &$\sim$0           \\
 $|(2S1s)3/2^+\rangle \to |(1P1s)3/2^- \rangle \gamma$      & 0.17       &0.17           &0.14      &$\sim$0      &$\sim$0         &$\sim$0           \\\hline
 $|(1S2s)1/2^+\rangle \to |(1P1s)1/2^- \rangle \gamma$      & $\sim$ 0    &$\sim$ 0       &$\sim$0      &$\sim$0      &$\sim$0         &$\sim$0           \\
 $|(1S2s)1/2^+\rangle \to |(1P1s)3/2^- \rangle \gamma$      & $\sim$ 0    &$\sim$ 0       &$\sim$ 0 &$\sim$0      &$\sim$0         &$\sim$0           \\
 $|(1S2s)3/2^+\rangle \to |(1P1s)1/2^- \rangle \gamma$      & 0.05        &0.05            &0.01      &$\sim$0      &$\sim$0         &$\sim$0           \\
 $|(1S2s)3/2^+\rangle \to |(1P1s)3/2^- \rangle \gamma$      & 0.41        &0.41            &0.35      &0.01      &0.01         &0.01           \\\hline
 $|(1S1p)1/2^-~_{S_T=1/2}\rangle \to |(1S1s)1/2^+ \rangle \gamma$      & 171.99   &142.88    &192.71      &296.48      &59.95         &56.07           \\
 $|(1S1p)3/2^-~_{S_T=1/2}\rangle \to |(1S1s)1/2^+ \rangle \gamma$      & 430.75   &233.47     &237.88      &531.16      &114.51         & 79.32          \\
 $|(1S1p)1/2^-~_{S_T=3/2}\rangle \to |(1S1s)1/2^+ \rangle \gamma$      & 27.66    &6.21       &0.93      &20.66      &5.20         &1.12           \\
 $|(1S1p)3/2^-~_{S_T=3/2}\rangle \to |(1S1s)1/2^+ \rangle \gamma$      & 126.26   &28.36      &4.01      &78.89      &19.84         &4.06           \\
 $|(1S1p)5/2^-\rangle \to |(1S1s)1/2^+ \rangle \gamma$                 & 105.45   &23.69      &3.26      &63.15      &15.88         &3.10           \\
 $|(1S1p)1/2^-~_{S_T=1/2}\rangle \to |(1S1s)3/2^+ \rangle \gamma$      & 56.29    &12.62       &1.97      &77.15      &19.41         &3.96           \\
 $|(1S1p)3/2^-~_{S_T=1/2}\rangle \to |(1S1s)3/2^+ \rangle \gamma$      & 71.96    &16.14      &2.67      &77.86     &19.58         &4.20           \\
 $|(1S1p)1/2^-~_{S_T=3/2}\rangle \to |(1S1s)3/2^+ \rangle \gamma$      & 403.02   &200.88     &183.53      &635.96      &141.35         &82.56           \\
 $|(1S1p)3/2^-~_{S_T=3/2}\rangle \to |(1S1s)3/2^+ \rangle \gamma$      & 394.72    &213.04     &207.93      &554.59      &121.00         &77.43           \\
 $|(1S1p)5/2^-\rangle \to |(1S1s)3/2^+ \rangle \gamma$                 & 31.30    &31.77        &65.55      &61.85      &11.70         &18.23        \\\hline

 $|(1S2s)1/2^+\rangle \to |(1S1p)1/2^-~_{S_T=1/2} \rangle \gamma$      & 46.42    &17.78          &24.00      &30.99      &7.13         &6.46           \\
 $|(1S2s)1/2^+\rangle \to |(1S1p)3/2^-~_{S_T=1/2} \rangle \gamma$      & 38.46    &28.60          &41.96      &44.78      &9.32         &9.05           \\
 $|(1S2s)1/2^+\rangle \to |(1S1p)1/2^-~_{S_T=3/2} \rangle \gamma$      & 3.09     &0.33           &0.04      &0.99        &0.28        &0.08   \\
 $|(1S2s)1/2^+\rangle \to |(1S1p)3/2^-~_{S_T=3/2} \rangle \gamma$      & 6.27     &0.66           &0.09      &2.82        &0.80        &0.27   \\
 $|(1S2s)1/2^+\rangle \to |(1S1p)5/2^- \rangle \gamma$                 & 0.34     &0.31           &0.05      &1.51        &0.43        &0.17   \\
 $|(1S2s)3/2^+\rangle \to |(1S1p)1/2^-~_{S_T=1/2} \rangle \gamma$      & 11.29    &1.19            &0.16      &2.77       &0.79        &0.27   \\
 $|(1S2s)3/2^+\rangle \to |(1S1p)3/2^-~_{S_T=1/2} \rangle \gamma$      & 8.72     &0.92            &0.11      &2.99       &0.85        &0.27   \\
 $|(1S2s)3/2^+\rangle \to |(1S1p)1/2^-~_{S_T=3/2} \rangle \gamma$      & 6.10     &2.99            &8.06      &7.65       &1.72       &2.31    \\
 $|(1S2s)3/2^+\rangle \to |(1S1p)3/2^-~_{S_T=3/2} \rangle \gamma$      & 20.16    &11.87            &18.83      &22.54    &4.88         &4.93           \\
 $|(1S2s)3/2^+\rangle \to |(1S1p)5/2^- \rangle \gamma$                 & 42.03    &27.90            &36.39      &46.55    &9.84         &8.62           \\
\hline\hline
\end{tabular}
\end{center}
\end{table*}

The radiative decays of the low-lying excited doubly heavy baryons are studied as well.
Our results are listed in Tab.~\ref{decay2}. It is found that for the
$(1P1s) \to (1S1s)\gamma$, $(1D1s) \to (1P1s)\gamma$, and $(2S1s) \to (1P1s)\gamma$
transitions, the partial radiative decay widths
of doubly charmed baryons range from several ten to hundred eV,
while the decays of doubly bottom baryons approximate vanish.
It should be mentioned that the pionic or kaonic strong decays
for $(1P1s)$, $(1D1s)$ and $(2S1s)$ containing diquark excitations are forbidden due to the orthogonality
of the diquark wave functions. Hence, the $(1P1s)$, $(1D1s)$ and $(2S1s)$ doubly heavy
baryons should be very narrow states, and their radiative transitions should play an
important role in their decays. These radiative transitions may be useful for looking
for the missing $(1P1s)$, $(1D1s)$ and $(2S1s)$ doubly charmed states.
Furthermore, it is found that most of the transitions
$(1S1p)\to (1S1s)\gamma$ have large partial decay widths, which can reach up to several
hundred keV. However, the radiative decay rates of $(1S1p)\to (1S1s)\gamma$ may
be seriously suppressed by the large decay widths of $(1S1p)$,
because the decays of $(1S1p)$ are dominated by the strong decay modes.
About the radiative transitions of the excited doubly heavy baryons,
few discussions are found in the literature, thus, more studies are expected
to be carried out in future.

\section{Summary}{\label{Summary}}

In this work, we study the mass spectra and radiative decays of doubly heavy baryons within the diquark picture in a relativized quark model. The diquark masses and wave functions are calculated from the relativized quark potential, and the doubly heavy baryon spectra are obtained by solving the Schr\"{o}dinger-type equation between the diquark and quark. The effects of the diquark sizes are
considered by introducing the form factors. Besides the ground doubly heavy baryons,
the masses for the low-lying excited doubly charmed and bottom baryons are given.
The theoretical mass of the ground state $\Xi_{cc}$ ($J^P=1/2^+$), 3606 MeV, is consistent with the mass
of the newly observed $\Xi_{cc}^{++}(3621)$ by the LHCb collaboration.
The predicted mass gap $m(\Xi^*_{cc})-m(\Xi_{cc})$ is about 69 MeV, which is waited to be tested in future experiments.
Furthermore, using the realistic wave functions obtained from the relativized quark model and the quark-photon couplings,
we evaluate the radiative transitions between these states. It is interesting to find that
the $(1P1s)$, $(1D1s)$ and $(2S1s)$ doubly charmed baryons containing diquark excitations
should be very narrow states with a width of several ten to hundred eV, the decays
of these states should be dominated by the radiative transitions due to the absence of the strong decay modes.
The radiative decay width of $\Xi_{cc}^{*++} \to \Xi_{cc}^{++}\gamma$ and $\Xi_{cc}^{*+} \to \Xi_{cc}^+\gamma$
are quite significant, which are up to about 7 and 4 keV, respectively.
We hope these theoretical predictions of doubly heavy baryons should be helpful for future experimental exploration.

\bigskip
\noindent
\begin{center}
{\bf ACKNOWLEDGEMENTS}\\

\end{center}
We would like to thank Yu-Bing Dong and Qiang Zhao for valuable discussions. This project is supported by
the National Natural Science Foundation of China under Grants No. 11705056 and No. 11375061.

\end{document}